\documentclass[journal=jacsat,manuscript=article]{achemso}
\usepackage{tgadventor}
\usepackage{achemso}
\usepackage{float}
\usepackage{graphicx}
\usepackage{amsfonts}
\usepackage{amsmath}
\usepackage{multirow}
\usepackage{wrapfig}
\usepackage[T1]{fontenc}
\usepackage{geometry}
\title{
On the adsorption of oxygen to high entropy alloy surfaces up to 2ML coverage using Density Functional Theory and Monte Carlo calculations}
\author{Tyler D. Dole\v{z}al}
\affiliation[AFIT]
{Department of Engineering Physics, Air Force Institute of Technology, 2950 Hobson Way, Wright-Patterson Air Force Base, OH 45433, USA}
\author{Adib J. Samin}
\affiliation{Department of Engineering Physics, Air Force Institute of Technology, 2950 Hobson Way, Wright-Patterson Air Force Base, OH 45433, USA}
\email{adib.samin@afit.edu}

\begin{document}
%%%%%%%%%%%%%%%%%%%%%%%%%%%%%%%%%%%%%%%%%%%%%%%%%%%%%%%%%
%%%%%%%%%%%%%%%%%%%%%%%%%%% HEADING %%%%%%%%%%%%%%%%%%%%%
%%%%%%%%%%%%%%%%%%%%%%%%%%%%%%%%%%%%%%%%%%%%%%%%%%%%%%%%%

%%%%%%%%%%%%%%%%%%%%%%%%%%%%%%%%%%%%%%%%%%%%%%%%%%%%%%%%%
%%%%%%%%%%%%%%%%%%%%%%%% ABSTRACT %%%%%%%%%%%%%%%%%%%%%%%
%%%%%%%%%%%%%%%%%%%%%%%%%%%%%%%%%%%%%%%%%%%%%%%%%%%%%%%%%
\begin{abstract}
    High entropy alloys (HEAs) manufactured with refractory elements are candidates for high temperature structural applications. To enhance our understanding of oxidation in these complex systems, the early stages of oxidation on the surface of Al${}_{10}$Nb${}_{15}$Ta${}_{5}$Ti${}_{30}$Zr${}_{40}$ were studied using Density Functional Theory and thermodynamic modeling. Surface slabs were generated from bulk configurations sampled in equilibrium using a multicell Monte Carlo method for phase prediction. The bulk structure was found to be a single BCC phase in good agreement with experimental observations. The oxygen adsorbed with a strong preference towards sites with Ti and Zr and avoided sites with Nb-Al and Nb-Ta. The surface was shown to be highly reactive to oxygen, yielding a dominating oxygen coverage of 2 monolayer over the temperature range of 100 to 2600 K and oxygen pressure range of $10^{-30}$ to $10^{5}$ bar. Recovering a clean surface slab was not achieved until pressures approached vacuum conditions and temperature exceeded 1900 K, demonstrating the difficulty of oxygen removal from the surface. Grand Canonical Monte Carlo simulations showed that high Nb content in the top surface layer reduced the surface reactivity to incoming oxygen. Inward oxygen diffusion at low coverage was preferred in regions rich with Zr, but slowed with the addition of Ti and Al. Diffusion rates drastically reduced at 1 ML, especially in the region rich with Ti and Zr, where strong metal-oxygen bonds were reported. Our results indicated that a high content of Ti and Zr increased the reactivity of the HEA surface to oxygen. The presence of Nb also enhanced resistance to oxygen adsorption, especially when partnered with Al and Ta. Inward oxygen diffusion was likely to occur at low coverage in regions rich with Zr, but could be protected against with the addition of Al and Ti. The limitations of the present work are discussed. This study may provide insights that assist with devising short- and long-term mitigation strategies against material degradation related to high temperature oxidation.       
\end{abstract}
\section*{Introduction}
High entropy alloys (HEAs) have been an area of intense research since their first mention in 2004, where Cantor et al., \cite{cantorMicrostructuralDevelopmentEquiatomic2004} synthesized a five component Fe${}_{20}$Cr${}_{20}$Mn${}_{20}$Ni${}_{20}$Co$_{20}$ alloy by melt spinning and Yeh et. al, \cite{yehNanostructuredHighEntropyAlloys2004} produced, and coined the term HEA, several HEAs by arc melting which included Cu, Ti, Cr, Ni, Co, V, Fe, and Al. They proposed that a multi-component alloy at near equiatomic concentrations would increase the entropy of mixing to overcome the enthalpies of compound formation, resulting in a lower probability of formation of potentially deleterious intermetallics. This statement is summarized mathematically by examining the ideal entropy of mixing,
\begin{equation}\label{eq:entropy}
    \Delta S_{mix} = -R\sum_{i=1}^n x_i\ln(x_i)
\end{equation}
where R is the ideal gas constant and $x_i$ is the atomic concentration of the i-th element. If n is to equal 4, 5, or 6, and all $x_i$ take on the same value, the value of $\Delta S_{mix}$ is 1.39R, 1.61R, 1.79R, respectively. The entropic contribution to the free energy at elevated temperatures, $T_h$, that is, -$T_h\Delta S_{mix}$, is on the same order as the enthalpy of formation for intermetallic compounds; thereby suppressing their probability of formation. Recent discussions suggest a more nuanced approach in the nomenclature of these new multi-component alloys \cite{miracleCriticalReviewHigh2017}.
\par 
Here, the focus is restricted refractory HEAs (RHEAs), which are promising candidate materials for high temperature structural applications, such as their use in jet propulsion systems as the structural material for the engine turbine blades and discs. Senkov and associates have done extensive experimental investigations on the mechanical properties, phase stability, and high temperature oxidation behavior of refractory alloy systems to examine the viability of deploying them in aircraft propulsion systems \cite{senkovMicrostructureMechanicalProperties2020,senkovTemperatureDependentDeformation2020,senkovMicrostructureMechanicalProperties2021,whitfieldAssessmentThermalStability2021, butlerOxidationBehaviorsCrNb2022}. Several RHEAs have been thoroughly discussed and recently reviewed by Couzinie et al. \cite{couzinieComprehensiveDataCompilation2018}. Experimental studies on the high temperature oxidation resistance of RHEAs and the role different metals play are on-going \cite{butlerOxidationBehaviorArc2016, mullerOxidationMechanismRefractory2019, huangRoleNbHigh2019, schellertOxidationMechanismRefractory2021,shiInfluenceMicroArc2021, yangEffectContentThermal2021}. Butler and Weaver \cite{butlerOxidationBehaviorArc2016} explored Al$_{x}$(
NiCoCrFe)$_{1-x}$ where the Al content was varied. They reported a combination of Al$_2$O$_3$ and AlN beneath an external Cr$_2$O$_3$ scale across all samples with the general conclusion that increasing Al content improved oxidation resistance by bolstering the position and continuity of the Al$_2$O$_3$ scale. M\"{u}ller et al., \cite{mullerOxidationMechanismRefractory2019} studied TaMoCrTiAl,
NbMoCrTiAl, NbMoCrAl and TaMoCrAl and reported the formation of a protective oxide layer containing Al$_2$O$_3$, Cr$_2$O$_3$ and CrTaO$_4$ in the quinary Ta-containing alloy with the general conclusion that increased Ta content reduced MoO$_3$ evaporation and increased oxidation resistance. Schellert et al., \cite{schellertOxidationMechanismRefractory2021} performed a similar analysis to \cite{mullerOxidationMechanismRefractory2019} on the same HEA where they found a Ta content < 10\% resulted in poor oxidation performance. In the computational domain, Taylor et al., \cite{taylorIntegratedComputationalMaterials2018} reported metal-oxide formation energies calculated with density functional theory (DFT) that suggested Zr, Ta, Nb, Ti, and Al are among the top candidates for oxide resistance with oxide formation energies exceeding 8 eV/atom.  Osei-Agyemang and Balasubramanian \cite{osei-agyemangSurfaceOxidationMechanism2019} employed DFT calculations and thermodynamic modeling to study the thermodynamics of oxygen adsorption on the surface of MoWTaTiZr which is among the first applications DFT calculations to study the oxidation of a RHEA surface. They found the surface of MoWTaTiZr to be highly reactive to oxygen with 1 ML oxygen coverage preferred at temperatures 300 K to 1500 K and were unable to recover a clean surface, even at extremely low pressures and high temperatures. 
\par\
This study examines the surface oxidation of Al${}_{10}$Nb${}_{15}$Ta${}_{5}$Ti${}_{30}$Zr${}_{40}$,\cite{soniPhaseStabilityMicrostructure2020} where oxygen coverage is increased from 0.07 monolayer to 2 monolayer (ML) on a surface slab generated from a bulk supercell \cite{saminFirstprinciplesInvestigationSurface2018,osei-agyemangSurfaceOxidationMechanism2019}.
Here, we use our implementation of the solid-phase phase prediction algorithm, (MC)${}^2$ \cite{niuMulticellMonteCarlo2019,antillonEfficientDeterminationSolidstate2020}, to predict the alloys bulk structure which is used to generate candidate surface structures. This is different from previous studies where the phase of the bulk material was assumed and the bulk alloy was generated randomly, rather than from equilibrium. DFT calculations and thermodynamic modeling were performed and stable oxygen coverage over a large scale of temperatures and pressures is examined. Oxygen was systematically adsorbed to the (011) surface and affinity towards the different metal atoms is discussed. Charge analysis was executed using the Bader code \cite{tangGridbasedBaderAnalysis2009}. Successive adsorption energy as a function of the surface composition and oxygen coverage was generated using the method of least squares. Four (011) surface slabs with differing metallic compositions were explored using the successive adsorption energy function in a Grand Canonical Monte Carlo Simulation (GCMC). Climbing image nudged elastic band (CI-NEB) calculations were employed to calculate the activation energies for five diffusion routes from the top surface layer to the first subsurface layer at low and 1 ML oxygen coverage. This work may help increase our knowledge as to how material properties are affected as the oxidation process occurs. Such properties are typically needed as input for continuum models \cite{saminFirstprinciplesInvestigationSurface2018}.
%%%%%%%%%%%%%%%%%%%%%%%%%%%%%%%%%%%%%%%%%%%%%%%%%%%%
%%%%%%%%%%%%% METHOD %%%%%%%%%%%%%%%%%%%%%%%%%%%%%%%
%%%%%%%%%%%%%%%%%%%%%%%%%%%%%%%%%%%%%%%%%%%%%%%%%%%%
\section*{Computational Methods}
%%%%%%%%%%%%%%% BULK %%%%%%%%%%%%%%%%%%%%%%%%
\subsection*{Bulk Structure}
The bulk structure was generated using our implementation of the (MC)${}^2$ algorithm \cite{niuMulticellMonteCarlo2019, antillonEfficientDeterminationSolidstate2020}. The simulation was performed at T = 1273.15 K, P = 0 Pa, and the five simulation cells were initialized with 32 atoms per cell in initial cubic structures of body centered cubic (BCC), hexagonal closed packed (HCP), face centered cubic (FCC), and BCC. Equilibrium was reached in about 350 steps and was sampled for an additional 1,311 steps for a total run consisting of 1,661 steps. During the simulation two moves were randomly selected, either a local flip or intraswap. A local flip is described as randomly selecting one of the five simulation cells, then randomly selecting one atom within the chosen simulation cell and "flipping" it from its current species type to one of the other four species types. The intraswap consists of randomly selecting one of the five simulation cells, then randomly selecting two atoms whose positions are swapped. The algorithm has been constructed to only perform an intraswap between two atoms of different types, should one simulation cell become 100\% of one species, an intraswap move is rejected. DFT calculations were executed to calculate simulation cell internal energy and volume changes; the settings are discussed below. The acceptance criteria, based on the Metropolis criteria \cite{metropolisEquationStateCalculations1953}, for the local flip and intraswap are given in Eq \ref{eq:accept-flip} and Eq \ref{eq:accept-swap}, respectively. 
\begin{equation}\label{eq:accept-flip}
P^{flip}_{accept} = min\left\{1,\exp\big(-\beta \Delta H + N\Delta G\big)\right\},
\end{equation}
where $\Delta H$ and $\Delta G$ are
\begin{gather}\label{eq:delta_g}
    \Delta H = m\sum_{i=1}^{m}(U'_{i} + pV'_{i})f'_{i} - m\sum_{i=1}^{m}(U_{i} + pV_{i})f_{i}\\
     \Delta G = \sum_{i=1}^{m}[f'_{i}\ln(V'_{i}) - f_{i}\ln(V_{i})] + \sum_{i=1}^{m}f'^{i}\sum_{j=1}^{m}X^{' i}_{j}\ln(X^{' i}_{j}) - \sum_{i=1}^{m}f^{i}\sum_{j=1}^{m}X^{i}_{j}\ln(X^{i}_{j}).
\end{gather}
Here, $\beta = 1/k_B T$, where $k_B$ is the Boltzmann constant, N is the sum of all the particles across all simulation cells, m is the total number of simulation cells, $U_i$ is the energy of simulation cell $i$, $V_i$ is the volume of simulation cell $i$, $p$ is the pressure, which was set to 0 Pa, and $f_i$ is the molar fraction of simulation cell $i$. Lastly, letting $n^i_j$ be the number of species $i$ in simulation cell $j$, $X^i_j = n^i_j/\sum_{k=1}^{5}n^k_j$, which is the atomic concentration of species $i$ in simulation cell $j$. The primed coordinates indicate post-flipped values, while un-primed are pre-flipped values.
\begin{equation}\label{eq:accept-swap}
P^{swap}_{accept} = min\left\{1,\exp(-\beta \Delta U_i)\right\},
\end{equation}
where $\Delta U_i$ is the change in the simulation cell's energy after the swap is performed.
\par
DFT calculations were performed using the Projector Augmented Wave (PAW) method as implemented by the Vienna ab initio Software Package (VASP) \cite{kresseEfficientIterativeSchemes1996,kresseUltrasoftPseudopotentialsProjector1999}. The calculations were performed with a plane wave cutoff energy of 450 eV and a 2x2x2 Monkhorst-Pack \cite{monkhorstSpecialPointsBrillouinzone1976} k-point mesh. DFT calculations performed on the simulation cells allowed for changes in the volume and atomic positions through the setting ISIF = 3. The electronic self-consistent calculation was converged to 1x10${}^{-6}$ eV and ionic relaxation steps were performed using the conjugate-gradient method (IBRION = 2) and continued until the total force on each atom dropped below a tolerance of 1x10${}^{-2}$ eV/\AA. The generalized gradient approximation (GGA) was used for the exchange correlation functionals as parameterized by Perdew-Burke and Ernzerof (PBE) \cite{perdewGeneralizedGradientApproximation1996}. The PAW pseudopotentials \cite{kresseUltrasoftPseudopotentialsProjector1999} were used with the valence electron configurations 3s${}^2$3p${}^1$, 4p${}^6$5s${}^1$4d${}^4$, 6s${}^2$5d${}^3$, 3d${}^3$4s${}^1$, and 5s${}^2$4d${}^1$5p${}^1$, for  Al, Nb, Ta, Ti, and Zr, respectively. 
%%%%%%%%%%%%%%%% SURFACE CALCS %%%%%%%%%%%%%%%%%%%%%%%
\subsection*{Surface Study}
\begin{figure}[H]
    \centering
    \includegraphics[width=0.6\linewidth]{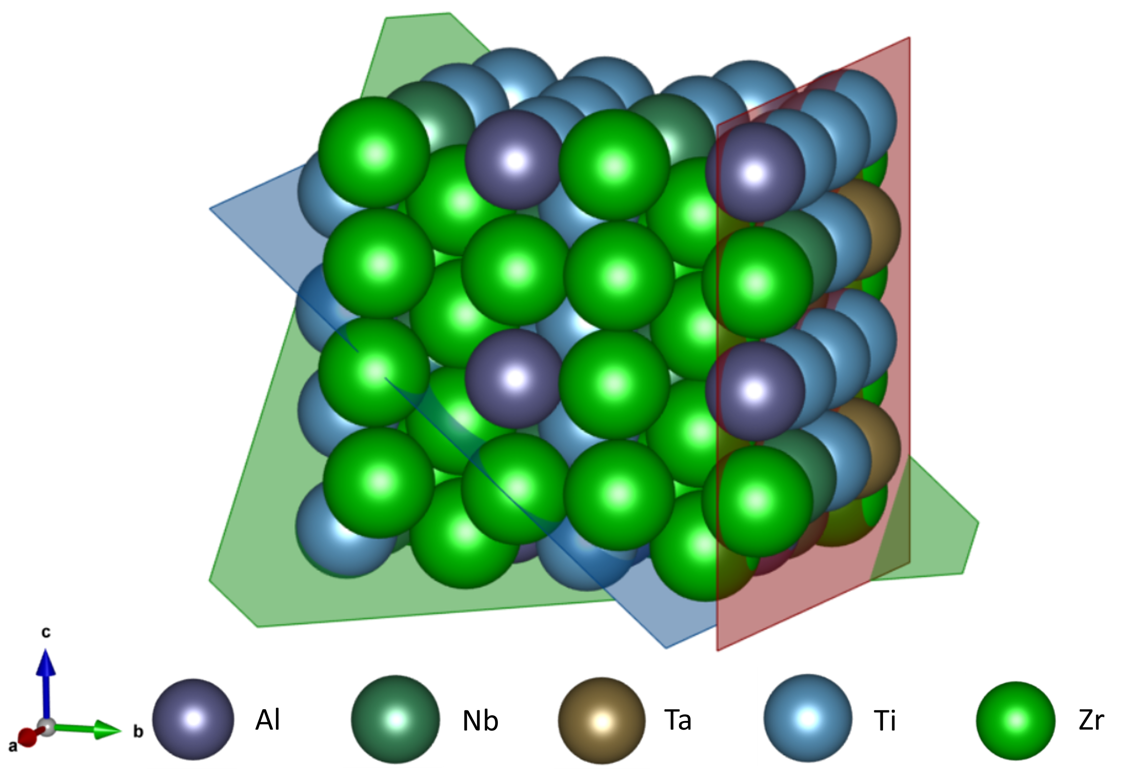}
    \caption{The body centered cubic bulk structure multiplied into a larger supercell for surface cuts. The (010), (011), and (111) surface planes are shown in red, blue, and green, respectively.}
    \label{fig:vesta-faces}
\end{figure}

In an effort to keep the number of computations reasonable and reduce the computational overhead, surface slabs were generated from the (MC)${}^2$ simulation cell with the largest molar fraction. In our case, this corresponded to the 4th simulation cell, a BCC crystal structure with atomic concentrations closely matching that of the RHEA BCC matrix. Initially, one equilibrium bulk configuration was extracted and surface slabs from the {100}, {110}, and {111} families were generated using our surface slab tool\cite{dolezalslabgenerator}. The bulk structure and surface cuts are shown in Figure \ref{fig:vesta-faces}. Surface terminations were chosen such that each slab contained all members of the alloy to maximize insight on how the adsorbed oxygen interacts with different members of the alloy. The surface slab with the lowest surface energy was chosen for additional study; for this work it was the (011) slab. Two additional bulk equilibrium configurations of the 4th simulation cell were extracted from (MC)${}^2$ and two new (011) surface slabs were generated. The (011) slab with the lowest surface energy was used for the adsorption study. The surface slabs consisted of four layers which were divided into two surface layers and two bulk layers. A vacuum layer of 20 $\textrm{\AA}$ was applied along the $\hat{c}$ direction to prohibit interaction between image slabs. Increasing the number of layers in our surface model was considered, however, after seeing a perpendicular response of less than 0.05 \AA\ by the second layer up to 2 ML oxygen coverage, more layers were not introduced. Selective dynamics was used to freeze the bulk layers while the surface layers and O-adsorbate(s) were free to move in the $\hat{a}$, $\hat{b}$, and $\hat{c}$ directions.
%----------------------------------- MC2 BULK FIGURE -------------------------------------------------
%------------------------------------------------------------------------------------------------------------
DFT calculations were executed with a plane wave cut-off energy of 600 eV and a 3x3x1 Gamma-point-centered Monkhorst-Pack k-point grid \cite{monkhorstSpecialPointsBrillouinzone1976}. The electronic energy was converged with respect to the k-point grid and energy cut-off to within 1 meV. Through propagation of error in the electronic energies, surface energies were reported with confidence up to 0.1 meV/\AA${}^2$. Electronic relaxation was converged to 1x10${}^{-6}$ eV and ionic relaxation steps were continued until a force tolerance criterion of 1x10${}^{-2}$ eV/\AA\ was satisfied. The PAW pseudopotentials had the same valence electron configurations as stated in the previous subsection, with the addition of an electron valence configuration of 2s${}^2$2p${}^4$ for oxygen. The surface energy per area is given by,
\begin{equation}\label{eq:surf-energy}
    E_{surf} = \frac{1}{2A}\left(E_{slab} - NE_{bulk}\right),
\end{equation}
where A is the area of the surface slab, in units \AA${}^2$, $E_{slab}$ is the electronic energy of the surface slab, in units of eV, N is the number of atoms present in the surface slab, and $E_{bulk}$ is the energy per atom of the bulk structure from which the surface slab was cut, in units of eV/atom.
\par 
Adsorption calculations were executed using an energy cut-off of 600 eV, an electronic relaxation convergence setting of 1x10${}^{-6}$ eV, and force tolerance criteria of 1x10${}^{-2}$ eV/\AA. Electronic energies were converged with respect to the k-point grid and plane wave cut-off energy to within 1 meV. Through propagation of error in the electronic energies, the adsorption energies were reported with confidence up to 1 meV. An attempt to activate dipole corrections along the $\hat{c}$ direction was made, but this resulted in poor convergence with the conjugate gradient method (IBRION = 2); therefore it was deactivated. Non-spin polarized calculations were performed after validating that spin-polarized calculations had a negligible effect (less than 0.05\%) on the adsorption energy of oxygen. The k-point grid was altered between 1x1x1 and 3x3x1 at different points in the routine and the projection operators were evaluated in real-space through the setting LREAL = Auto, with a final, static, calculation being performed with the projection operators done in reciprocal space. Making coarse measurements (1x1x1, LREAL=Auto) that were later refined (3x3x1, LREAL=.FALSE.) greatly reduced the computational cost of this work. Adsorption of the O$_2$ molecule showed a preference for dissociated oxygen adsorption, in agreement with the findings of similar metal-oxygen adsorption studies \cite{erikatDensityFunctionalStudy2011,xingAdsorptionDiffusionOxygen2021}. Therefore, we consecutively adsorbed oxygen in the dissociated form. To limit the number of adsorption sites to consider, 6 adsorption calculations were performed with a single O atom placed at 2 different on-top, bridge, and hollow sites using a k-point grid of 3x3x1. The O atom was placed 1.75 \AA\ above the surface for the on-top and bridge sites and 1.25 \AA\ above the surface for the hollow sites. The 2 on-top and bridge calculations resulted in the O atom displacing to a hollow site and both hollow-placed O atoms remained in the hollow site they were placed. Therefore, only hollow sites were considered when consecutively placing O atoms to increase the coverage.
\par
For every additional O atom placed, two rounds of adsorption calculations were made on the unoccupied hollow sites. The first round was conducted using a k-point grid of 1x1x1. From the first round, the three hollow sites with the lowest adsorption energies were selected and the second round of adsorption calculations was executed with a 3x3x1 k-point grid for a more accurate energy. The O atom was placed at the site with the lowest adsorption energy. When three and fewer hollow sites remained unoccupied the routine would execute only the 3x3x1 adsorption calculation and choose the minimum from that point. The adsorption and successive adsorption energies are given by,
\begin{equation}\label{eq:ads-energy}
    E_{ads} = \left(E^{O(n)}_{slab} - E_{slab} - \frac{n}{2}E_{O_2}\right)
\end{equation}
\begin{equation}\label{eq:succ-ads-energy}
    E^{Succ}_{ads} = \left(E^{O(n+1)}_{slab} - E^{O(n)}_{slab} - \frac{1}{2}E_{O_2}\right)
\end{equation}
where $E^{O(n)}_{slab}$ is the total electronic energy of the oxidized slab, $E_{slab}$ is the total electronic energy of the clean slab, $n$ is the number of O atoms present, $E_{O_2}$ is the energy of the gas phase O${}_2$ molecule, and $E^{O(n+1)}_{slab} - E^{O(n)}_{slab}$ is the difference in electronic energy between the surface with $n$ O-atoms and $n+1$ O-atoms present. The electronic energy of O${}_2$ was calculated by placing two O atoms in a box of dimensions (8, 8, 8) \AA\ separated along the $\hat{c}$ direction by 1.22 \AA. The electronic energy was converged with respect to box size. The calculated O$_2$ electronic energy was -9.86 eV. The surface slab coverage was increased from 1/15 monolayer (ML) to 15/15 ML using a step of 1/15 ML. Once full coverage was achieved, a 16th O atom was introduced to the fully covered surface slab. The 16th O atom was restricted to move only in the $\hat{c}$ direction to avoid adsorption at top layer bridge sites. The 16th adsorption event occurred at the site where the first O atom was adsorbed and caused the O atom which already occupied this site to move in the -$\hat{c}$ direction to the first subsurface layer. The original O atom diffused below the hollow site it previously occupied, on-top of Zr, surrounded by 3 Ti and 1 Zr. This event had a successive adsorption energy of $E^{Succ}_{ads}$ = -2.61 eV, a value larger than the largest first-layer adsorption event by 1.52 eV. This positive difference in adsorption energy demonstrates that an O-O repulsion event which results in O diffusing to the subsurface will not occur until all adsorption sites are occupied. The monolayer study was resumed from the 15/15 ML surface structure, where O atoms were introduced to hollow adsorption sites on the first subsurface layer. The same routine to progress from 1/15 to 15/15 ML was performed to progress from 16/15 to 30/15 ML coverage at a step of 1/15 ML. Once finished, the work function was calculated for the clean slab up to the 30/15 ML oxygen coverage.  
\par
The Bader algorithm \cite{tangGridbasedBaderAnalysis2009} was used to examine the charge transfer between the surface slab and O-atoms. Using the clean surface slab as the point of reference, the charge transferred is defined as,
\begin{equation}\label{eq:bader-charge}
    \Delta q_j = q_{i,0} - q_{i,j},
\end{equation}
where $j$ indexes the ML structure, and $i$ indexes the alloy atoms in the surface slab, not to include the oxygen, and $q_{i,0}$ is the charge value of atom $i$ in the clean surface. The result is a matrix of charge transfer which catalogues the difference of each atom's charge at each ML structure in reference to its initial charge value. As it is defined in Eq. \ref{eq:bader-charge}, a positive $\Delta q$ value indicates a gain of electrons.
\par 
The Gibbs free energy of formation per atom was calculated for the coverage configurations using the following expression,
\begin{equation}\label{eq:gibbs-energy}
    \Delta G = \left((E_{slab}^{O(n)} + F_{vib}) - E_{slab} - n\mu_{O}(T,P) - TS_{mix}\right)/(N+n)
\end{equation}
where $E^{O(n)}_{slab}$ is the total electronic energy of the oxidized slab, $E_{slab}$ is the total electronic energy of the clean slab, $N$ is the number of metal atoms, and $n$ is the number of O atoms present. $F_{vib}$ refers to the Helmholtz vibrational energy and is a function of temperature and phonon frequency. The expression for the Helmholtz vibrational energy, within the harmonic approximation, is given as,
\begin{equation}
    F_{vib} = -k_B T \ln(Z) = \frac{1}{2}\sum^{3n}_{\textbf{q}j}\hbar \omega_j(\textbf{q}) + k_B T\sum^{3n}_{\textbf{q}j} \ln\left[1-\exp(-\hbar\omega_j(\textbf{q})/k_B T)\right],  
\end{equation}
where Z is the vibrational partition function, and the number of normal modes, or frequencies, is equal to the number of O atoms present in the system, $n$, multiplied by the number of degrees of freedom, i.e. $3n$ normal modes. The phonon frequencies for point $\vec{q}$ in the first Brillouin Zone and band index $j$, $\omega_j(\textbf{q})$, were calculated using the finite-difference method as implemented by VASP (IBIRON = 5) with $\Gamma$-point sampling only. The term $\mu_O$ represents the chemical potential of oxygen and is expressed as,
\begin{equation}\label{eq:chem-pot}
    \mu_{O}(T,P_{O_2}) = \frac{1}{2}E_{O_2} + \Delta \mu_O(T,P_{O_2}). 
\end{equation}
The first term in Eq. \ref{eq:chem-pot} is the energy of the gas phase O$_2$ molecule. $\Delta \mu_{O}$ is a correction term that is treated as a parameter and is expressed as,
\begin{equation}\label{eq:chem-corr}
    \Delta \mu_O(T,P_{O_2}) = \frac{1}{2}\left(\mu^0_{O_2}+k_B T\ln\left(\frac{P_{O_2}}{P^{0}}\right)\right).
\end{equation}
The term $\mu^0_{O_2}$ in Eq. \ref{eq:chem-corr} is the difference of the chemical potential of O$_2$ at T = 0 K and T > 0 K under P = 1 atm and can be calculated at different temperatures while holding the pressure constant using the values reported for a range of temperatures in the National Institute of Standards and Technology (NIST) Joint Army-Navy-Air Force (JANAF) Thermochemical tables as demonstrated by Cleary and Samin\cite{zotero-251,clearyUseJANAFTables2014,saminOxidationThermodynamicsNbTi2021}. The entropy of mixing term, $S_{mix}$, in Eq. \ref{eq:gibbs-energy} is defined in Eq. \ref{eq:entropy} and is a constant value in this case because the atomic concentrations of each species, $x_i$, remained unchanged.
\begin{equation}\label{eq:entropy}
    S_{mix} = -k_B\sum_{i=1}^{5} x_i\ln(x_i)
\end{equation}
For this study, $P^0$ was set to 1 bar and the system was analyzed in the temperature and pressure ranges of [100 K, 2600 K] and [$10^{-30}$ bar, $10^{5}$ bar], respectively. For each set of (T,P) the coverage with the lowest Gibbs free energy was identified and a stability plot was produced. The temperature bin width for the stability plot is 100 K and was restricted to that of the JANAF tables. The pressure range was divided into 35,000 data points; a finer division was tested, but led to negligible improvements. For a given temperature, the Gibbs free energy was calculated for each of the 35,000 pressures. Then, the list of energies was scanned and the coverage with the lowest Gibbs energy at each pressure value was recorded.
\par
Using the surface adsorption study results a least squares generated function was derived for the successive adsorption energy as a function of the number of metal and oxygen atoms present within a cutoff radius of 3.5 \AA\ of the adsorption site; this included first and second surface layer atoms at some sites. A linear function form was assumed,
$$E^{Succ}_{ads}(\textbf{n},n_{O}) = a n_{Al} + b n_{Nb} + c n_{Ta} + d n_{Ti} + e n_{Zr} + f n_{O}$$
where a, b, c, d, e, f are coefficients found using the method of least squares, \textbf{n} is the number of each metal atom in the slab, and $n_{O}$ is the number of O atoms present on the slab. The coefficient values and a parity plot are included in the supplemental document\footnote{Additionally, the reader can find a comparison between alternative methods for generating a linearly dependent adsorption energy function in the supplemental document}. The matrix elements, $a_{i,j}$, represent the number of metal atom type $j$ present at adsorption site $i$. The coefficients were stored in a coefficient vector, \textbf{w}. For different terminations of the (011) slab an \textbf{A} matrix can be generated and operated on \textbf{w} to yield the successive adsorption energies expected on that slab. The successive adsorption energies for three new configurations of the original (011) slab were used in a Grand Canonical Monte Carlo (GCMC) simulation. Two moves were attempted, oxygen insertion and oxygen deletion, whose acceptance probabilities are given in Eq. \ref{eq:gcmc-insert} and \ref{eq:gcmc-delete}.
\begin{equation}\label{eq:gcmc-insert}
    P^{insert}_{accept} = min\left\{1,\frac{V}{\lambda^3(N+1)}\exp(\beta \mu)\exp(-\beta \Delta U)\right\}
\end{equation}
\begin{equation}\label{eq:gcmc-delete}
    P^{delete}_{accept} = min\left\{1,\frac{\lambda^3 N}{V}\exp(-\beta \mu)\exp(-\beta \Delta U)\right\}
\end{equation}
In Eqs. \ref{eq:gcmc-insert} and \ref{eq:gcmc-delete}, V is the simulation cell volume, $\lambda$ is the thermal de Broglie wavelength, N is the number of atoms present, metals and oxygen, $\beta$ = 1/$k_B$T where $k_B$ is the Boltzmann constant, T is temperature, $\mu$ is the oxygen chemical potential given in Eq. \ref{eq:chem-pot}, and $\Delta U$ is the energy difference which is given in Eq. \ref{eq:gcmc-energy},
\begin{equation}\label{eq:gcmc-energy}
    \Delta U = E^{Succ}_{ads} +\frac{1}{2}E_{O_2}.
\end{equation}
The temperature domain was swept from 100 K to 2600 K with a step of 100 K, the pressure domain was swept from $10^{-30}$ bar to $10^{5}$ bar with a step of $10^{3}$ bar and a stability plot was generated up to 1 ML. Each GCMC simulation was executed for 5,000 steps. To examine the impact of each species' contribution to the surface resistance against oxygen adsorption, three new slabs were generated where 8 Zr atoms on either the first or second surface layer were flipped to either Al, Nb, or Ti. Because Ta was scarce on the training slab, fluctuations in Ta content was not explored.
\par
\begin{figure}[H]
    \centering
    \includegraphics[width=0.65\linewidth]{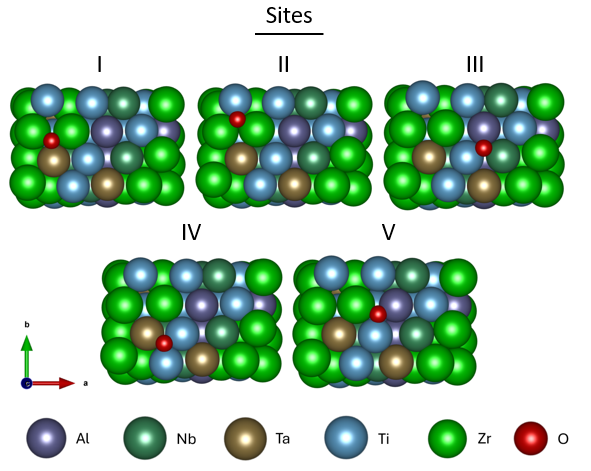}
    \caption{The initial slab configuration for the five diffusion events}
    \label{fig:NEB}
\end{figure}

 
The energy barrier for an O atom to diffuse from the top surface layer to the first subsurface layer was calculated using the climbing image nudged elastic band (CI-NEB) method \cite{henkelmanClimbingImageNudged2000} as implemented by the software VASP Transition State Theory (VTST). The calculation was carried out using six images and a spring constant of - 5.0 eV/\AA${}^2$. The initial and final positions were completely relaxed before calling the CI-NEB calculations. The initial configuration for each of the five sites is shown, as seen in VESTA\cite{mommaVESTAThreedimensionalVisualization2011}, in Figure \ref{fig:NEB}. Sites I, II, and III were also analyzed at 1 ML coverage to examine diffusion at high coverage. The convergence parameters remained unchanged from the previous VASP calculations and a k-point density of 3x3x1 was used.
%------------------------------------ NEB SITE FIGURE ------------------------------------------------
%-----------------------------------------------------------------------------------------------------
To examine the time scale over which the diffusion events would occur, the calculated activation energies were used in the Arrhenius rate equation, as shown in Eq. \ref{eq:reax-rate}. The rate constant for each event was inverted to yield an order of magnitude estimate for the time associated with it.
\begin{equation}\label{eq:reax-rate}
    k_{i} = \nu_0\exp\left(E_{i}/k_B T\right)
\end{equation}
In Eq. \ref{eq:reax-rate}, $k_i$ is the rate constant of event $i$, in units Hz, $\nu_0$ was set to the commonly used value of 10$^{12}$ Hz for solid-state diffusion, and $E_i$ is the energy corresponding to event $i$, in units eV.
%%%%%%%%%%%%%%%%%%%%%%%%%%%%%%%%%%%%%%%%%%%%%%%%%%%%%%%%%
%%%%%%%%%%%%%%%%%%%% DISCUSSION %%%%%%%%%%%%%%%%%%%%%%%%%
%%%%%%%%%%%%%%%%%%%%%%%%%%%%%%%%%%%%%%%%%%%%%%%%%%%%%%%%%
\section*{Results \& Discussion}
\begin{figure}[H]
    \centering
    \includegraphics[width=\linewidth]{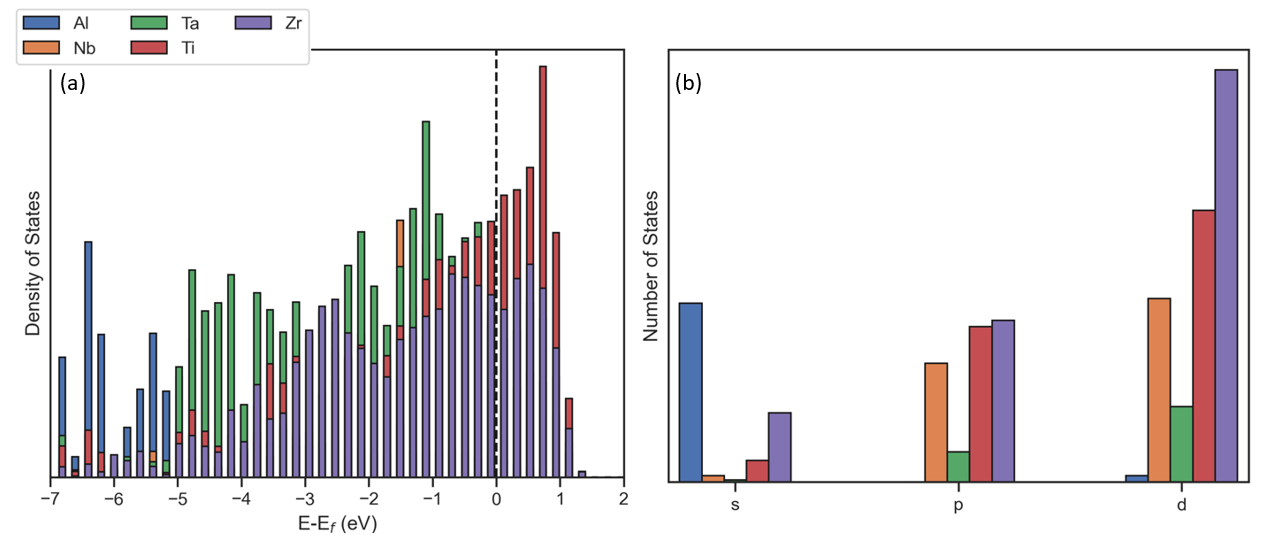}
    \caption{(a) The projected density of states for the bulk structure. (b) State distribution over the s-, p-, and d-orbitals}
    \label{fig:PDOS}
\end{figure}

\begin{figure}[H]
    \centering
    \includegraphics[width=0.55\linewidth]{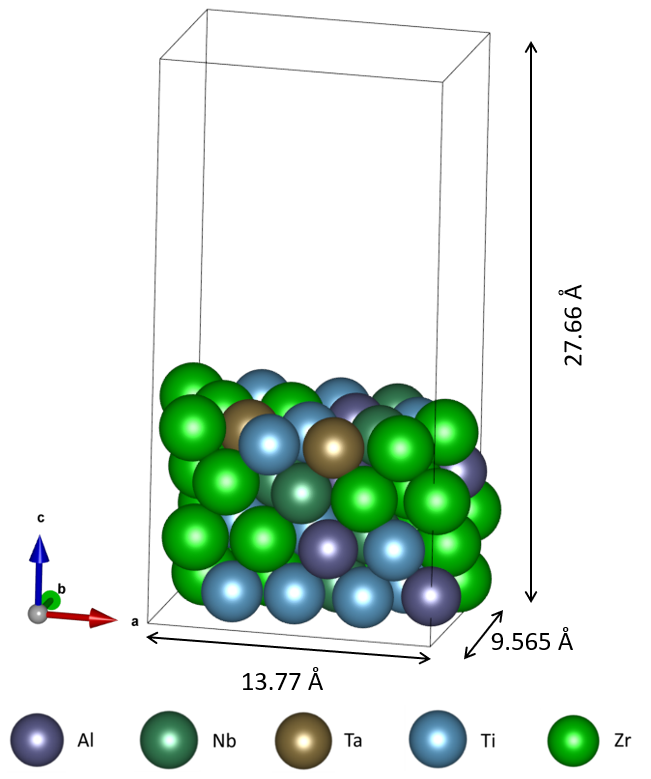}
    \caption{The (011) surface slab generated from the body centered cubic bulk structure}
    \label{fig:b2-surface}
\end{figure}

(MC)${}^2$ predicted a majority phase with a molar fraction of 87\% at the composition of Al${}_{9}$Nb${}_{16}$Ta${}_{3}$Ti${}_{31}$Zr${}_{41}$ by atomic percent (at.\%) and Zr${}_{50}$Ti${}_{20}$Nb${}_{20}$Ta${}_{7}$Al${}_{3}$ by weight percent (w.\%), at 1273.15 K, P = 0 bar. This result was in good agreement with the experimentally observed single BCC phase seen at T = 1373.15 K. The fidelity of the at.\% values are 0.03125, or 1/32. To reduce this value, and therefore increase the precision of the predictions, more atoms need to be included in the simulation cells. When using a software like VASP, doing so greatly increases the cost of the simulation.
The average lattice constant of the bulk structure was 3.402 \AA\ which is in excellent agreement with the experimentally measured lattice constant of 3.401 \AA. Figure \ref{fig:PDOS} (a) shows the projected density of states (DOS) for the bulk structure used for surface slab generation. Figure \ref{fig:PDOS} (b) is a histogram which shows each species distribution of states over the s-, p-, and d-orbitals. Ti and Zr have a high number of d-orbital states in the conduction band and near the Fermi level which resulted in a strong interaction between O and these species. 
The surface energy for the (010), (011), and (111) were 122.0, 110.6, and 121.3 meV/\AA$^2$, respectively. The additional (011) surface slab energies varied by about 1 meV/\AA$^2$. The (011) surface slab used in the oxygen adsorption study is shown, as seen in VESTA, in Figure \ref{fig:b2-surface} and had the lowest surface energy out of the three (011) configurations.
%%%%%%%%%%%%%%%%%%%%%%%%%%%%%%%%%%%%%%%%%%%%%%%%%%%%%%%%%%%%%%%%%%%%%%%%%%%%%%%%%%%%%%%%%%%
%%%%%%%%%%%%%%%%%%%%%%%%%%%%%%%%%% END OF MC2 DISCUSSION %%%%%%%%%%%%%%%%%%%%%%%%%%%%%%%%%%
%%%%%%%%%%%%%%%%%%%%%%%%%%%%%%%%%%%%%%%%%%%%%%%%%%%%%%%%%%%%%%%%%%%%%%%%%%%%%%%%%%%%%%%%%%%
%------------------------------------- ADS O ATOM ORDER ------------------------
\begin{figure}[H]
    \centering
    \includegraphics[width=\linewidth]{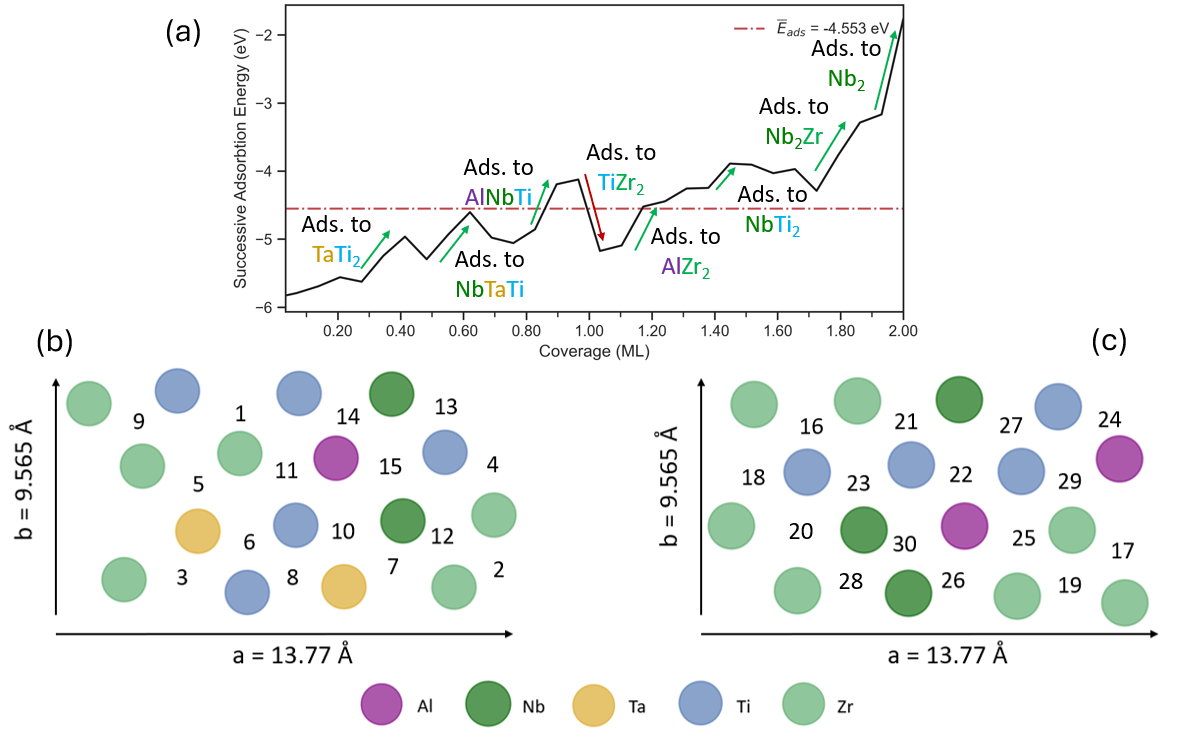}
    \caption{(a) Successive adsorption energies plotted against the coverage with some notable events labeled. (b) The top-most layer of the surface slab with each adsorption event labeled in order from 1st to 15th oxygen atom adsorbed. (c) The second surface layer with each adsorption event labeled in order from 16th to 30th oxygen atom adsorbed.}
    \label{fig:succ-ads-energy}
\end{figure}
%----------------------------------------------------------------
Figure \ref{fig:succ-ads-energy} (a) shows the successive adsorption energies versus coverage, (b) is the top surface layer with the order of oxygen adsorption labeled, and (c) is the second surface layer with the order of oxygen adsorption labeled. The successive adsorption energy varied as sites with different metallic combinations were explored and as the local surface geometry changed in response to adsorption events. Sites with Ti and Zr present corresponded to a decrease in adsorption energy on both surface layers, indicating that the presence of Ti and Zr increased the likelihood of oxygen adsorption to that site. Adsorption to sites with Nb present resulted in an increase in adsorption energy, especially in the absence of Ti and Zr. Adsorption to a site where the combination of Nb/Al and Nb/Ta was present resulted in a large increase in adsorption energy. The general upward trend of successive adsorption energy is attributed to increased O-O repulsion as neighboring sites became occupied. A strong case of this was seen at the 30th adsorption event, where O-O repulsion forces the final oxygen to settle at a bridge adsorption site, which in turn pushes a neighboring O atom atop Zr. Referring to Figure \ref{fig:PDOS}, preferential adsoprtion by O to sites with Ti and Zr is explained by the strong covalent bonds formed as a result of the unfilled d-orbitals of Zr and Ti. In a similar theoretical investigation on the oxidation resistance of HEAs, Osei-Agyemang and Balasubramanian also reported the preferential adsorption of oxygen to sites with Ti and Zr\cite{osei-agyemangSurfaceOxidationMechanism2019}. Adsorption energy per adsorbate steadily increased from -5.8 eV/O to -4.5 eV/O as oxygen coverage reached 2 ML. It is likely that oxygen adsorption remains favorable beyond 2 ML oxygen coverage. While the adsorption energies agree with similar DFT-based surface adsorption studies\cite{guoTantalumSurfaceOxidation2017,saminFirstPrinciplesInvestigation2017, saminFirstprinciplesInvestigationSurface2018,osei-agyemangSurfaceOxidationMechanism2019}, the reported values for some oxides are about 2 eV greater than known oxide bond energies, NbO, TiO, and ZrO, for example.\cite{loockIonizationPotentialsBond1998} This comes from the propagation of error in the poorly predicted energy value for the O$_2$ molecule, where we reported -9.86 eV and calculated a binding energy of -6.45 eV, where $BE = E_{O_2} - 2E_{O}$, which is overestimated when compared to the known value of -5.12 eV. The molecular and atomic oxygen energy value varies depending on which exchange correlation functional is used \cite{droghettiPredictingTextdMagnetism2008,klupfelEffectPerdewZungerSelfinteraction2012}. In this work, the poor prediction of the O$_2$ molecule energy is due to the use of the PBE exchange-correlation functional, since DFT calculations which used the PBE functional have been shown to perform poorly at generating accurate surface adsorption energies\cite{ruizDensityfunctionalTheoryScreened2016}.
%---------------------------------- SLAB PROGRESSION ----------------------------------
\begin{figure}[H]
    \centering
    \includegraphics[width=\linewidth]{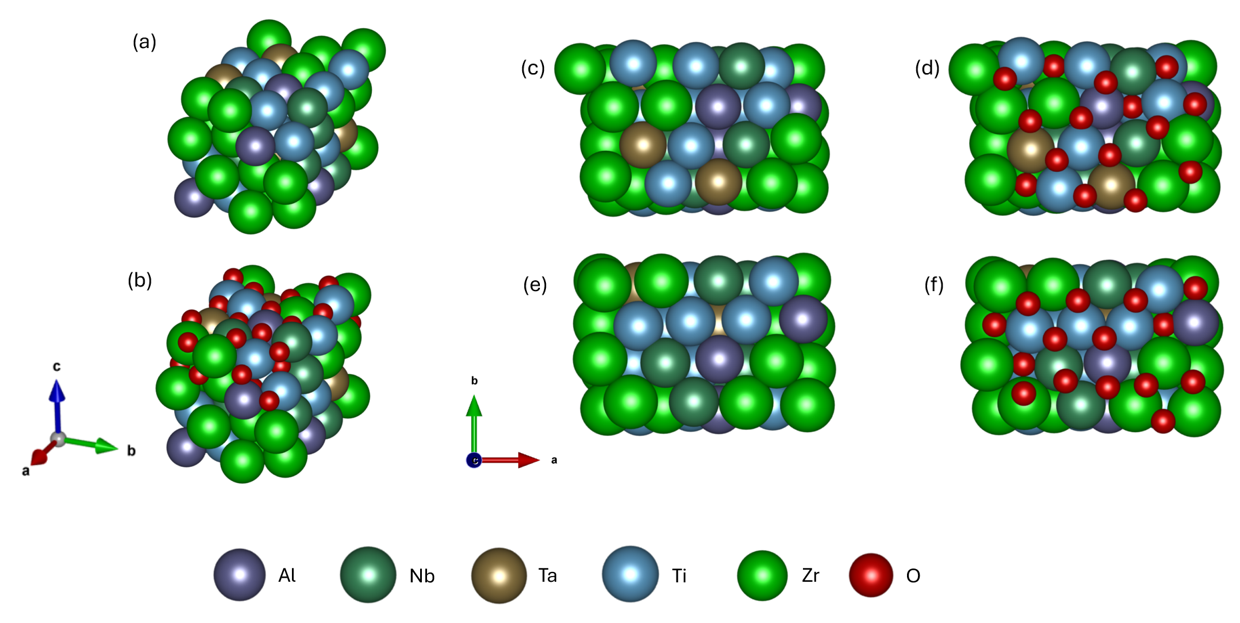}
    \caption{(a) Full view of the initial (011) surface slab. (b) Full view of the surface slab at 2 monolayer. (c) The first surface layer with no oxygen present. (d) the first surface layer at 1 monolayer. (e) the second surface layer with no oxygen present. (f) the second surface layer at 2 monolayer.}
    \label{fig:slab-progression}
\end{figure}
%----------------------------------------------------------------------------------------------------
The final configuration of the surface slab is shown in Figure \ref{fig:slab-progression}, where (a) and (b) shows the initial and final surface structure, (c) and (d) shows the top surface layer at 0 ML and 1 ML, and (e) and (f) shows the second surface layer at 0 ML and 2 ML. The distance between the first and second surface layer increased by 0.5 \AA\ from 0 ML to 2 ML while the distance between the second surface layer and first bulk layer saw an increase less than 0.05 \AA. Second layer surface atom positions were largely unperturbed as oxygen coverage on the top surface layer was increased. With respect to the surface plane, the average parallel and perpendicular movement of the second surface from 0 ML to 1 ML was less than 0.05 \AA. Increasing the oxygen coverage from 1 ML to 2 ML led to inward movement of 2 top surface Ta atoms, and a top surface Zr and Ti atom. 
\par
The work function was calculated using the expression, $\phi = E_{vac}-E_{F}$, where $E_{vac}$ is found from the averaged electrostatic potential and $E_{F}$ is the Fermi energy. The work function decreased until 0.27 ML where it remained fairly constant until a large increase around 0.80 ML. A comparison between the structure at 0.80 ML, 0.93 ML, and 1.00 ML showed a large displacement of Al by the 15th O atom, pushing Zr and Al together. The change in surface structure is a good candidate behind the boost in work function. The work function was relatively constant from 1.00 ML to 2.00 ML. The difference in the work function between clean and 2.00 ML was 0.118 eV. An increase in work function can be explained by the change in electrostatic dipole moment \cite{wuInteractionElectronTransfer1988} as charge is transferred from the metal surface to the O atoms; the shift of the surface is to a more positive state, while the O atoms become more negative, and this influences the work function. A decrease in the work function value indicates the alloy has a weaker bind on its electrons, making oxygen binding more thermodynamically favorable. Hugosson et al., \cite{hugossonSulfurOxygenInducedAlterations2013} reported an increase in the work function as oxygen coverage increased on the (001) surface of iron. Huber and Kirk \cite{huberWorkFunctionChanges1966} observed the reduction of the Al work function as oxygen coverage increased. A DFT-based surface adsorption study was conducted by Guo et al. on the pure Ta surface where they reported negligible changes in the work function vs. oxygen coverage on the (110) face\cite{guoTantalumSurfaceOxidation2017}. The mixture of five different metals on the surface leads to a more complicated behavior in the work function. 
\par
The M-O bond lengths at the first and second surface layer were calculated and compared. The M-O bond length was shorter at the second layer, indicating the formation of stronger bonds. When comparing the M-O bonds at low oxygen coverage against high oxygen coverage it was found that bond length decreased with increasing O population for all metal types.
%----------------------- BADER CHARGE FIGURE -------------------------
\begin{figure}[H]
    \centering
    \includegraphics[width=\linewidth]{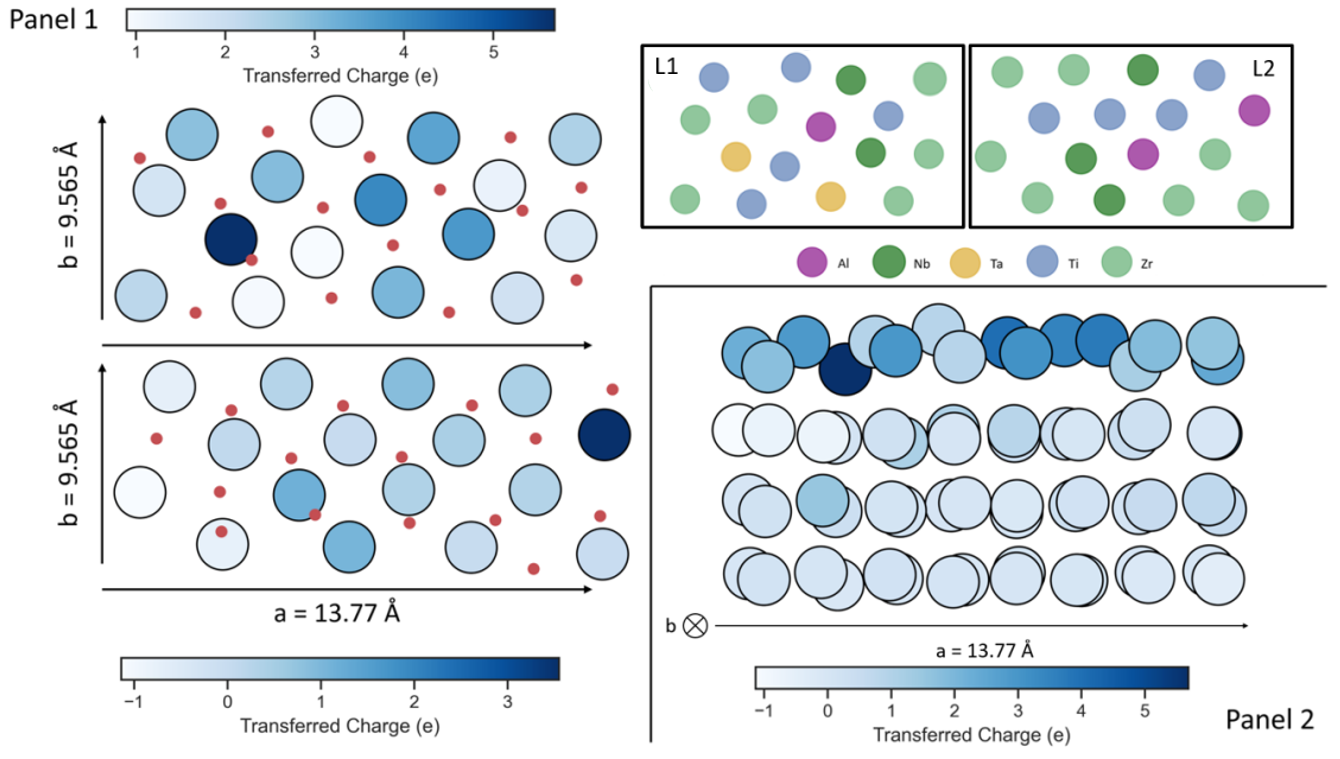}
    \caption{(Panel 1) The charge transferred by members of surface layer 1, (Top Left), and surface layer 2, (Bottom Left), at 1 and 2 monolayer coverage. Two site maps of surface layer 1 and 2 have been included in the upper right corner of Panel 1 to help readers identify which members occupy which positions. (Panel 2) A side view, where b points into the page, of the surface slab showing how charge is transferred as a function of depth.}
    \label{fig:Bader-charge}
\end{figure}
%----------------------------------------------------------------------------------------------------
The results from the Bader charge analysis are shown in Figure \ref{fig:Bader-charge}, where the first and second surface layer, referred to as L1 and L2 respectively, are displayed under full coverage, 1.00 ML and 2.00 ML, in (Panel 1). To help identify which alloy species occupies which site in the tranferred charge plots, a site map of L1 and L2 has been included in the upper right corner of (Panel 1). Figure \ref{fig:Bader-charge} (Panel 2) shows the transferred charge between the four layers of the surface slab. Beginning with (Panel 1), the strongest transfer of charge to the O atom in L1 is from the Ta and Al, followed by Nb, Ti, and Zr. The charge transferred values reported for Ta are consistent with the findings of a first-principles investigation on the oxidation of the pure Ta surface.\cite{guoTantalumSurfaceOxidation2017} In L2, the highest transfer was from Al followed by Nb, Ti, and Zr. Note the difference in the L1 and L2 heat map values, the strongest charge interaction occurs between the O atoms and L1. This is further supported in the layer-by-layer comparison shown in (Panel 2) where the highest amount of charge, darkest color, is seen in L1. An experimental investigation into the oxidation resistance of several HEAs at T = 1373 K yielded the conclusion that the presence of Ta reduced the evaporation of MoO$_3$ in the alloy TaMoCrAl\cite{mullerOxidationMechanismRefractory2019}. A similar result on Ta was reported by Schellert et al., \cite{schellertOxidationMechanismRefractory2021} with the addition that Ta content < 10\% failed to enhance the oxidation performance of TaMoCrAl. It could be that, when present in high enough amounts, the Ta-O charge interaction prohibits the formation of the MoO$_3$ covalent bonds\cite{yuFirstPrinciplesStudyMo2015}, thus weakening the oxygen-assisted outward movement of Mo which leads to the eventual evaporation of MoO$_3$. A similar analysis can be applied to Nb. An experimental investigation into the oxidation resistance of CoCrFeMnNb$_{0.25}$Ni at T = 973 K, 1073 K, and 1173 K showed Nb discouraged outward diffusion of metal cations \cite{huangRoleNbHigh2019}. It stands to reason that the surface behavior observed here, where Nb became more positive as oxygen adsorbed around it, could explain how Nb near the surface repels positively charged metal ions diffusing outward. The large charge transferred value of the L1 Ta and L2 Al is likely an artifact of the Bader algorithm's placement of Bader volumes when dealing with atoms of varying sizes. Aside from the two cases of extreme values, the magnitude of charge appears in good agreement to those reported by another ab initio HEA investigation \cite{jiangElasticThermodynamicProperties2020}.
%--------------------------- STABILITY PLOT ---------------------------------------------------
\begin{figure}[H]
    \centering
    \includegraphics[width=\linewidth]{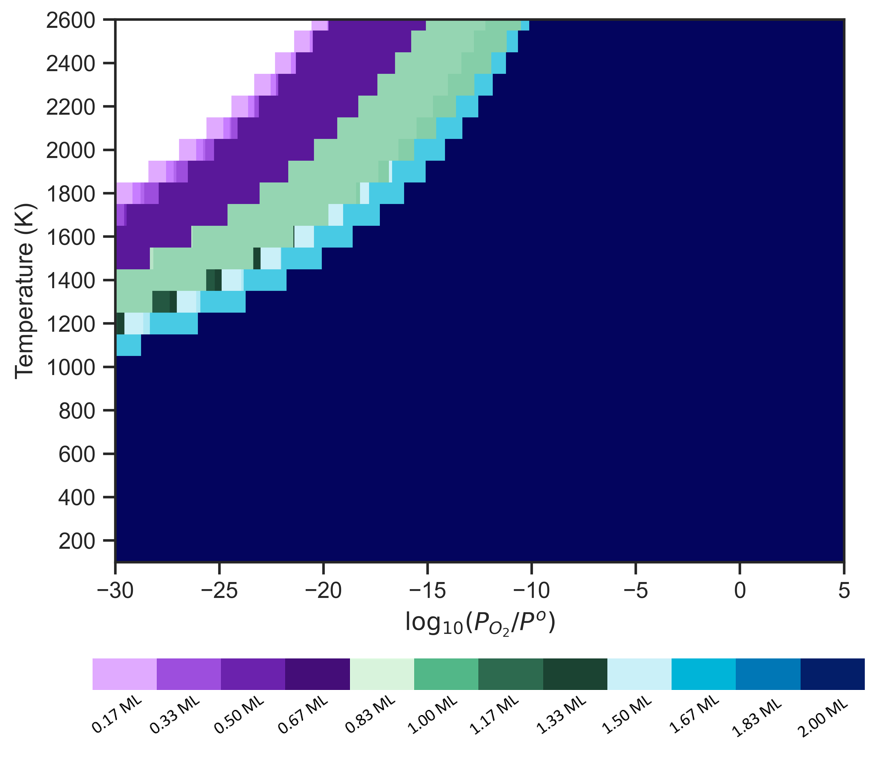}
    \caption{The stability plot which was generated using the results of the thermodynamics study. Extreme temperatures and pressures were considered only for academic purposes. The shading has been assigned to indicate increasing coverage, where white represents the clean surface, purples represent low coverage, greens represent moderate coverage, and blues represents high coverage.}
    \label{fig:stability-plot}
\end{figure}
%----------------------------------------------------------------------------------------
Figure \ref{fig:stability-plot} is the key output from the thermodynamics study. Removal of oxygen from the (011) surface of Al${}_{9}$Nb${}_{16}$Ta${}_{3}$Ti${}_{31}$Zr${}_{41}$ was not possible under realistic conditions and 2.00 ML is the thermodynamically stable oxygen coverage over the domain of operational possibilities. A similar theoretical result was reported by Osei-Agyemang and Balasubramanian \cite{osei-agyemangSurfaceOxidationMechanism2019} where a systematic investigation of oxygen adsorption on a HEA surface up to 1 ML revealed a strong affinity for oxygen, even at pressures as low as $10^{-9}$ bar and temperatures as high as 2000 K. The investigation to such extreme pressures was purely academic as we wished to identify the set of (T,P) values where the clean slab is preferred. A clean (011) surface was recovered at P $<$ $1\times 10^{-18}$ bar and T $>$ 1900 K. From a thermodynamic point of view, this strong affinity for high oxygen content brings into question the alloy's performance in long-term applications at elevated temperatures and under service conditions. One such long-term application could be the deployment of this alloy to an aircraft system. As of 2013, the typical aircraft lifespan is between 20 and 36 years\cite{jiangKeyFindingsAirplane2013}. The ageing of an aircraft is complex and costly\cite{dixonMaintenanceCostsAging2006} and depends on advancement in science and technology to reinforce the technical ageing of these systems. Another long-term application could be the deployment of this alloy in a very-high-temperature reactor (VHTR) system\cite{streetVeryHighTemperature}. Structural materials deployed in said system would need to withstand extended periods of time at high temperatures in corrosive environments.  
\par
The GCMC simulations reported negligible change in the surface reactivity to oxygen with increased amounts of Al and Ti. An increase in Nb content at the top surface slightly improved the surface resistance to oxygen adsorption. This result cannot be generalized to say increasing Nb increased the oxidation resistance. Huang et al. examined the CoCrFeMnNb$_0.25$Ni and found that, in the bulk, Nb forms a Laves phase which promoted inward diffusion of oxygen ions \cite{huangRoleNbHigh2019}. Jia et al. experimentally examined Ni-25Cr-10Fe-3Si-$\chi$Nb, up to 3 wt.\% Nb, and reported a reduction in oxidation resistance with increasing Nb content. The authors found that Nb would form Nb$_2$O$_5$ which reduced spalling resistance of the oxide scale\cite{jiaOxidationBehaviorNi25Cr10Fe3SihNb2021}.   
\par
\begin{figure}[H]
    \centering
    \includegraphics[width=\linewidth]{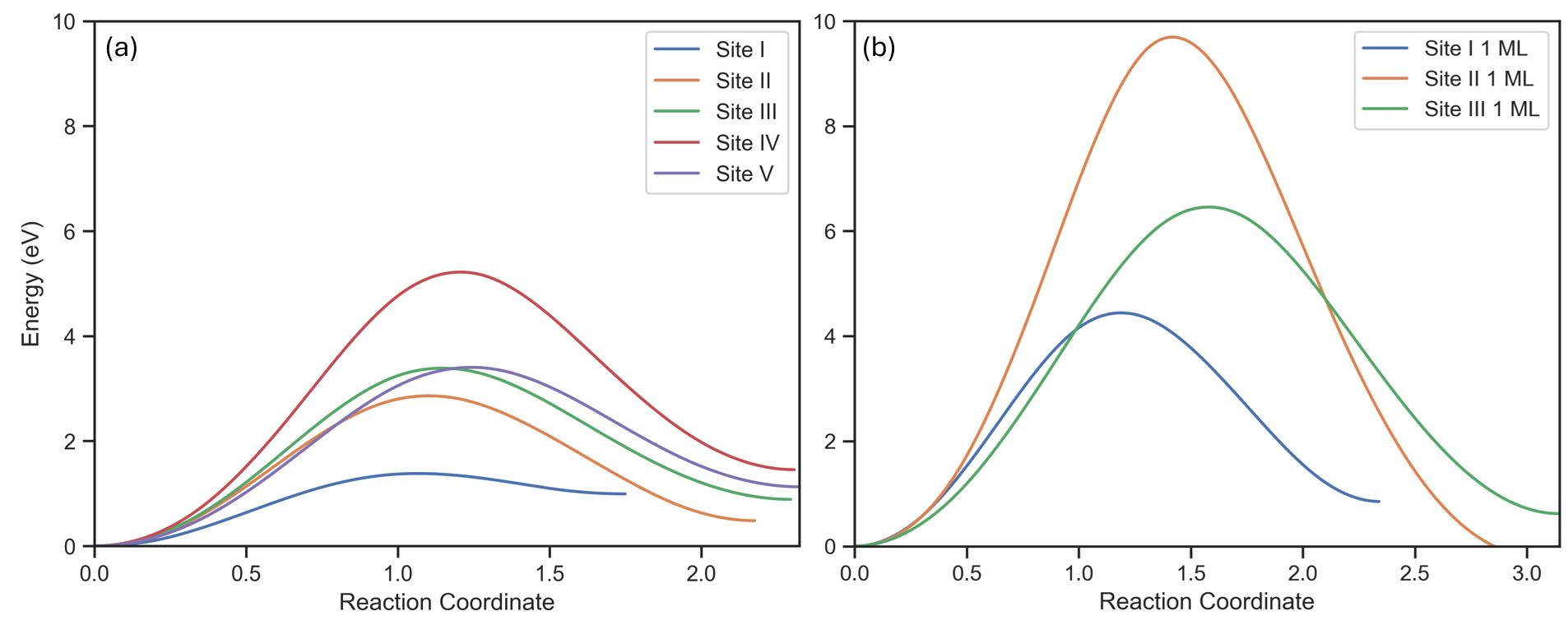}
    \caption{(a) The reaction pathway for the five diffusion sites at low coverage. (b) The reaction pathway for Site I, II, III at 1 monolayer coverage.}
    \label{fig:neb-reax}
\end{figure}

Table \ref{tab:neb-results} shows the results of the CI-NEB study and Figure \ref{fig:neb-reax} shows the energetic pathway plotted against the reaction coordinate, movement in the -$\hat{c}$ direction in this case. The \textbf{Metal} category in Table \ref{tab:neb-results} lists what metal atoms were present at the intitial site. Site I investigated the diffusion from a top layer hollow site with 2 Zr, 1 Ta to a second layer pseudo-stable bridge site between a Zr and Ti. Site II was the diffusion from a top layer hollow site with 2 Zr and 1 Ti to a second layer hollow site with 2 Zr and 1 Ti. Site III was the diffusion from a top layer hollow site with 1 Al, 1 Ti, and 1 Nb to a second layer hollow site with 2 Ti and 1 Al. Site IV was the diffusion from a top layer hollow site with 2 Ti and 1 Ta to a second layer hollow site with 1 Zr and 2 Nb. Site V was the diffusion from a top layer hollow site with 1 Al, 1 Ti, and 1 Zr to a second layer hollow site with 1 Al, 1 Nb, and 1 Ti. Sites where Zr was present had the lowest activation energies. Comparing Site I against Site II and V, diffusion at Zr-sites was slowed in the presence of Ti and Al. In fact, the highest single-oxygen diffusion route was at Site IV, which has the highest Ti content. Sites with Al present exhibited higher activation energies, as well.
%----------------------------- NEB RESULTS TABLE -----------------------
\begin{table}[H]
    \centering
    \caption{Results from the CI-NEB caclulations. The time is the inverted rate constant to its order of magnitude and the metal category lists the three nearest neighbors of the adsorbed oxygen}
    \begin{tabular}{cccccc}\hline\hline
    & & \multicolumn{3}{c}{\textbf{Time (s)}} & \\
    \textbf{Site}&\textbf{Activation Energy (eV)}&\textbf{1073 K} & \textbf{1173 K} & \textbf{1273 K} & \textbf{Metal}\\\hline
    \begin{tabular}{@{}c@{}}\textbf{I} \\\textbf{1ML} \end{tabular} & \begin{tabular}{@{}c@{}} 1.38 \\ 4.44 \end{tabular} & \begin{tabular}{@{}c@{}} $10^{-6}$ \\ $10^{8}$ \end{tabular} & \begin{tabular}{@{}c@{}} $10^{-6}$ \\ $10^{7}$ \end{tabular}& \begin{tabular}{@{}c@{}} $10^{-7}$ \\ $10^{5}$ \end{tabular}& \begin{tabular}{@{}c@{}} Ta, Zr$_{2}$\end{tabular}
    \\\hline
    \begin{tabular}{@{}c@{}}\textbf{II} \\\textbf{1ML} \end{tabular} & \begin{tabular}{@{}c@{}}2.86 \\ 9.69 \end{tabular} & \begin{tabular}{@{}c@{}} $10^{1}$ \\ $10^{33}$ \end{tabular} & \begin{tabular}{@{}c@{}} $10^{0}$ \\ $10^{29}$ \end{tabular}&\begin{tabular}{@{}c@{}} $10^{-1}$ \\ $10^{26}$ \end{tabular} & \begin{tabular}{@{}c@{}} Ti, Zr$_{2}$\end{tabular}
 \\\hline
    \begin{tabular}{@{}c@{}}\textbf{III} \\\textbf{1ML} \end{tabular} & \begin{tabular}{@{}c@{}}3.39 \\ 6.46 \end{tabular} & \begin{tabular}{@{}c@{}}$10^{4}$\\ $10^{18}$ \end{tabular} & \begin{tabular}{@{}c@{}} $10^{2}$ \\ $10^{15}$ \end{tabular}&\begin{tabular}{@{}c@{}} $10^{1}$ \\ $10^{12}$ \end{tabular} & \begin{tabular}{@{}c@{}} Al, Nb, Ti\end{tabular}
    \\\hline
    \textbf{IV} & 5.22 & $10^{12}$ & $10^{10}$ & $10^{8}$& Ta, Ti$_2$\\
    \textbf{V} & 3.40 & $10^{4}$ & $10^{2}$ & $10^{1}$& Al, Ti, Zr\\\hline\hline
    \end{tabular}
    \label{tab:neb-results}
\end{table}
%------------------------------ REAX PATHWAY PLOT ----------------------------------
%--------------------------------------------------------------------------------
Increasing oxygen coverage to 1 ML reduced the diffusion rates at Sites I, II, III. The increase in activation energy seems to be correlated to the increased rigidity of the top surface layer once multiple oxygen atoms are present and tightly bound to the metal atoms. At Site I, the second surface bridge site the O diffused to at low coverage was no longer accessible. At low coverage, a top layer Zr moved in response to the Site I O atom diffusion. At 1 ML, that Zr is bound to two additional O atoms and does not reveal the bridge site. The strong M-O bonds were also responsible for the large increase in activation at Site II, where the neighboring 2 Zr and 1 Ti are tightly bound to 2 additional O atoms. Similar results were seen with Site 3. The preliminary conclusion is that diffusion inward is slowed as coverage increases, where strong M-O bonds prohibit large structural changes that would allow inward migration of O. The time category of Table \ref{tab:neb-results} offers insight on the timescale for the diffusion events as predicted by the Arrhenius rate expression. Site I is the only diffusion event at 1 ML that occurs within a human lifespan, moving from tens of years to a few days as temperature increased. Oxygen diffusion at Site I and II, which had the highest concentration of Zr, occurred on the shortest time scales for all temperatures considered. Oxygen diffusion at Site IV, where Ti content was highest, has the highest event times at low oxygen coverage. While we must refrain from making any generalizations, the results on kinetics suggest that inward diffusion is heavily facilitated in regions rich with Zr. The diffusion times at sites with Zr were shown to slow with the addition of Ti and Al. The time scale at low oxygen coverage spanned from a fraction of a microsecond to days and years, depending on the temperature and local metallic environment.  
\par 
An experimental investigation into the oxidation resistance of Zr at 1273 K yielded a surprising result which showed Zr oxidation resistance increased with the formation of an oxide layer\cite{parkEnhancementOxidationResistance2018}. The authors found that the oxide layer promoted the formation of a large single crystalline ZrO grain which prohibited further oxidation. Here, we saw a strong affinity by the incoming oxygen to regions of the slab rich in Zr and Ti and strong resistance to oxygen inward diffusion at high oxygen coverage by sites containing Zr. The strong Zr-O interactions we report here, as well as, the increased resistance to inward diffusion at 1 ML oxygen coverage, could begin to explain how the ZrO grain observed by Park et al. prohibits further oxidation of the Zr alloy. Yang and Han \cite{yangEnhancedOxidationResistance2020}, examined the effects of surface implantation of Ti$^+$ on the oxidation resistance of Nb at 873 K. The authors found that the presence of Ti$^+$ on the surface resulted in a TiO$_2$ oxide scale and a reduction in the inward spread of oxygen into pure Nb. The strong Ti-O interactions and favorable adsorption energies beyond single oxygen adsorption to Ti sites suggest a formation of this oxide scale is possible. From the CI-NEB calculations we found that Ti bolstered against inward oxygen diffusion which agrees with the experimental observations of Yang and Han.  
\par
The findings of this work are consistent with experimental observations and may help provide insight about the mechanisms behind these observations from a thermodynamic and kinetic point of view. The results of our work may assist in understanding and predicting the complex oxidation behavior of RHEAs. However, no scientific study is without limitations. (MC)${}^{2}$ was executed with simulation cells consisting of 32 atoms, restricting the fidelity at which we scanned the compositional domain of the structure and limiting our ability to comment on the large-scale structure of the material. Smaller cells are also more susceptible to finite size effects. Surface analysis was performed on one (011) surface structure. Behaviors reported here should not be generalized to what may occur on the \{100\} and \{111\} family of surfaces. A more thorough investigation should include results observed on representations from the three surface families. Introducing the oxygen adsorbates systematically to the first layer followed by the second layer did not account for diffusion of oxygen to the second surface layer which we did show are likely to occur at low oxygen coverage. Vibrational corrections were applied to oxygen atoms only and under the the harmonic approximation. Including the phonon contributions of the metal atoms and anharmonic corrections were not considered and may have played an important role under certain conditions. The method used to generate the successive adsorption energy function is rudimentary when compared to previously reported models \cite{saminFirstPrinciplesInvestigation2017,saminCombinedDensityFunctional2018}, as these account for first and second nearest neighbor interactions, as well as, three and four body interactions. While the scanning of the (T,P) domain was insightful, it should be made clear that we did not consider phase transitions, which were experimentally observed\cite{soniPhaseStabilityMicrostructure2020}, of the bulk structure, and therefore the surface slab geometry, as we scanned the domain. The most accurate values of the stability plot relate to the predictions made near 1273 K, where the bulk structure and thermodynamically favored surface are valid. The kinetics of oxygen diffusion was not thoroughly examined here. A more comprehensive study would examine diffusion inward from a multitude of different routes with different metal atoms present at the initial and final position and at different oxygen coverage. Additionally, we did not account for thermal expansion which is likely to occur at high temperatures. It is reasonable to assume that the activation energies would change as a result of expansion.
%%%%%%%%%%%%%%%%%%%%%%%%%%%%%%%%%%%%%%%%%%%%%%%%%%%%%%%%%
%%%%%%%%%%%%%%%%%% Conclusion %%%%%%%%%%%%%%%%
%%%%%%%%%%%%%%%%%%%%%%%%%%%%%%%%%%%%%%%%%%%%%%%%%%%%%%%%%
\section*{Conclusion}
This work examined several aspects of the Al${}_{10}$Nb${}_{15}$Ta${}_{5}$Ti${}_{30}$Zr${}_{40}$ RHEA. First, we determined a single-phase BCC structure existed at 1,000 $^o$C with a lattice constant of 3.402 \AA\ which is in good agreement with experiment. Oxygen strongly favored adsorption sites where Zr and Ti were present and avoided sites where Nb was present. The work function increased with increasing oxygen coverage. M-O bond length decreased as coverage increased, which increased the rigidity of the two surface layers. Charge transferred to the oxygen was most prominent at the top surface layer where Ta, Al, and Nb were the strongest donors in the alloy. The (011) surface was found to be highly reactive with oxygen, with 2 ML coverage dominating the majority of the (T,P) domain studied in the stability plot. The GCMC simulation showed that increasing the Nb content slightly reduced the surface reactivity to oxygen, while increased Al and Ti content had negligible effects. Diffusion to the second surface layer preferred regions rich in Zr, but was slowed with the addition Ti and Al. Inward diffusion at 1 ML coverage was drastically reduced, especially in a region rich with Ti and Zr, which could be due to the strong metal-oxygen bonds that formed at an oxygen coverage of 1 ML. This system requires further investigation before results can be generalized on how each metal species impacts the early-stage oxidation resistance. However, the results discussed within could provide important insight into the early-stage mechanisms of oxidation-related structural degradation of HEAs and their viability as long-term structural candidates in systems which operate at high temperatures.
%%%%%%%%%%%%%%%%%%%%%%%%%%%%%%%%%%%%%%%%%%%%%%%%%%%%%%%%%
%%%%%%%%%%%%%%%%%% DATA AVAILABILITY %%%%%%%%%%%%%%%%
%%%%%%%%%%%%%%%%%%%%%%%%%%%%%%%%%%%%%%%%%%%%%%%%%%%%%%%%%
\section*{Data Availability}
The (MC)${}^2$ algorithm that was used to generate the bulk structure, all data generated from this work, a script that automates o-adsorption, the tool we developed for cutting surface slabs, and the post-processing scripts are available on GitHub at,\\
\textbf{(MC)}$^2$, https://github.com/SaminGroup/Dolezal-MC2\\
\textbf{Generating (MC)$^2$ sim cells}, https://github.com/SaminGroup/Dolezal-MC2\_Simcells\\
\textbf{Data from the study}, https://github.com/SaminGroup/Dolezal-Surface-Calcs\\ 
\textbf{Surface slab tool}, https://github.com/SaminGroup/Dolezal-SlabGenerator
%%%%%%%%%%%%%%%%%%%%%%%%%%%%%%%%%%%%%%%%%%%%%%%%%%%%%%%%%
%%%%%%%%%%%%%%%%%% ACKNOWLEDEMENTS %%%%%%%%%%%%%%%%
%%%%%%%%%%%%%%%%%%%%%%%%%%%%%%%%%%%%%%%%%%%%%%%%%%%%%%%%%
\section{Supporting Information}
Supporting information was provided with this manuscript and includes a discussion on 1) the linear least squares models that were considered in this work, 2) the (MC)$^2$ algorithm, 3) the automated simulation cells script, 4) the surface slab tool script, and 5) the computational cost breakdown of the research.
\section{Acknowledgements}
A.J. Samin would like to acknowledge funding from the Air Force Office of Scientific Research (AFOSR) FY22 Small Grants program grant number: FA9550-19-S-003. In addition, the work was supported by computational allocations from the department of defense high performance computing through AFRL HPC Mustang.
%%%%%%%%%%%%%%%%%%%%%%%%%%%%%%%%%%%%%%%%%%%%%%%%%%%%%%%%%
%%%%%%%%%%%%%%%%%% AUTHOR CONTRIBS %%%%%%%%%%%%%%%%
%%%%%%%%%%%%%%%%%%%%%%%%%%%%%%%%%%%%%%%%%%%%%%%%%%%%%%%%%
\section{Author Contributions}
\textbf{Tyler Dole\v{z}al}: Performed the calculations, generated the code required for the study, performed data acquisition 
and analysis, prepared and edited the manuscript\\
\textbf{Adib J. Samin}: Review and editing, supervision, funding acquisition.\\
Both authors contributed to the methodology and conceptualization of the study.
%%%%%%%%%%%%%%%%%%%%%%%%%%%%%%%%%%%%%%%%%%%%%%%%%%%%%%%%%
%%%%%%%%%%%%%%%%%% Computational Details %%%%%%%%%%%%%%%%
%%%%%%%%%%%%%%%%%%%%%%%%%%%%%%%%%%%%%%%%%%%%%%%%%%%%%%%%%

%%%%%%%%%%%%%%%%%%%%%%%%%%%%%%%%%%%%%%%%%%%%%%%%%%%%%%%%%
%%%%%%%%%%%%%%%%%% BIBLIOGRAPHY %%%%%%%%%%%%%%%%%%%%%%%%%
%%%%%%%%%%%%%%%%%%%%%%%%%%%%%%%%%%%%%%%%%%%%%%%%%%%%%%%%%
\bibliographystyle{chem-acs}
%\bibliography{main}

\providecommand{\latin}[1]{#1}
\makeatletter
\providecommand{\doi}
  {\begingroup\let\do\@makeother\dospecials
  \catcode`\{=1 \catcode`\}=2 \doi@aux}
\providecommand{\doi@aux}[1]{\endgroup\texttt{#1}}
\makeatother
\providecommand*\mcitethebibliography{\thebibliography}
\csname @ifundefined\endcsname{endmcitethebibliography}
  {\let\endmcitethebibliography\endthebibliography}{}

\end{document}